**Ultrahigh-Entropy Compositionally Complex Ceramics: Fluorite–Pyrochlore Phase Stability and Order–Disorder Transitions**

Keqi Song [a, b], Jian Luo [a,b,*]

[a] Aiiso Yufeng Li Family Department of Chemical and Nano Engineering, University of California San Diego, La Jolla, California 92093, U.S.A.

[b] Program in Materials Science and Engineering, University of California San Diego, La Jolla, California 92093, U.S.A.

## Abstract

Three groups of 32 ultrahigh-entropy compositionally complex ceramics (CCCs) containing 16-19 components were synthesized and characterized. The first group of 15 CCCs, $(15RE_{1/15})_2TM_2O_7$ ($15RE_{1/15}$ = equimolar mixture of 15 rare-earth elements; TM = Ce, Zr, Hf, Ti, or their equimolar combinations), forms single ultrahigh-entropy fluorite or pyrochlore phases in 12 compositions, while two compositions exhibit fluorite–fluorite dual phases, observed for the first time. Compared with ternary $A_2B_2O_7$ (fluorite–pyrochlore transition at $r_A/r_B \approx 1.46$), these ultrahigh-entropy CCCs are more prone to ordering (pyrochlore formation), surprisingly even more at higher size disorder (more lattice distortion), with suppressed dual-phase formation relative to five-component high-entropy ceramics. Two new 19-component series, $(15RE_{1/15})_2[(Ce_{1/3}Zr_{1/3}Hf_{1/3})_yTi_{1-y}]_2O_7$ ($0 \leq y \leq 1$; denoted as "19CCC-P|F$_y$") and $(15RE_{1/15})_{2-z}(CeZrHf)_{(1+z)/2}Ti_{(1-z)/2}O_{8-\delta}$ ($0 \leq z \leq 1$; denoted as "19CCC-P|F$_z$"), both form single ultrahigh-entropy fluorite or pyrochlore phases. The 19CCC-P|F$_y$ series, maintaining a 2:2 ratio of 3+ and 4+ cations, exhibits a pyrochlore-to-fluorite order-disorder transition (ODT) at $y \approx 0.84$ (critical $\overline{r_A}/\overline{r_B} \approx 1.40$), consistent with the linear projection of the pyrochlore superstructure peak intensity as an order parameter. In contrast, the 19CCC-P|F$_z$ series with non-2:2 ratios of 3+ and 4+ cations show an abrupt ODT at $z \approx 0.41$-$0.44$, despite a projected transition at $z \approx 0.64$ from the order parameter.

[*] Corresponding Author. Email: jluo@alum.mit.edu (Jian Luo)

## 1. Introduction

High-entropy ceramics (HECs), commonly defined as multicomponent ceramic materials with an ideal configurational mixing entropy of $\Delta S_{mix} > 1.5k_B$ per cation on at least one sublattice, have attracted considerable research attention since 2015 [1,2]. In 2020, Luo and co-workers extended the concept of HECs to encompass a broader class of compositionally complex ceramics (CCCs), ranging from medium-entropy ceramics (MECs, $\Delta S_{mix} \approx 1{\sim}1.5k_B$ per cation) to ultrahigh-entropy ceramics (UECs, $\Delta S_{mix} > 2.3k_B$) [1,3]. Within this expanded framework, non-equimolar design as well as long- and short- range cation ordering has been employed to tailor material properties. In some cases, medium-entropy compositions have been shown to outperform their higher-entropy counterparts , suggesting that maximizing configurational entropy alone is not necessarily the most effective strategy for further optimizing material properties. Nevertheless, UECs offer a vast compositional design space, within which new phase-stability and phase-transition phenomena have been observed in several studies, as discussed later. These emerging physical phenomena may provide new opportunities for discovering and designing CCCs with previously unexplored structures and properties. In addition to single-phase CCCs, recent studies have demonstrated that thermal properties can be tuned by tailoring phase fractions in dual-phase CCCs, leading to the proposal of a corresponding dual-phase CCC design strategy [4,5].

Fluorite-based oxides demonstrate ultralow thermal conductivities ($k$) with significant applications in thermal barrier coatings (TBCs) [6–8]. Fluorite-based oxides encompass a broad family of crystal structures derived from the cubic fluorite phase ($AO_{2-\delta}$ or disordered defect fluorite $A_4O_7$, $Fm\bar{3}m$, No.225). Among these, the ordered cubic pyrochlore phase ($A_2B_2O_7$, $Fd\bar{3}m$, No. 227) has attracted the most attention. It can be viewed as a $2 \times 2 \times 2$ superstructure of the disordered defect fluorite basis structure, and the pyrochlore and fluorite structures are directly connected to each other through an order-disorder transition (ODT) of the cations and anions. In addition, another ordered derivative structure, the orthorhombic weberite phase (of $2 \times \sqrt{2} \times \sqrt{2}$ superstructure), can also form in certain compositions.

Considerable efforts have been devoted to developing descriptors for predicting phase formation in fluorite-based oxides. The cation-radius ratio $r_A/r_B$ ratio has proven particularly effective. For ternary $A_2B_2O_7$ zirconates and hafnates, a critical cation-radius ratio of $r_A/r_B \approx 1.46$ defines the transition threshold between fluorite and pyrochlore structures. When $r_A/r_B$ is below this threshold, the rare-earth cations and Zr or Hf share a single cation site, stabilizing the disordered fluorite structure. In contrast, when $r_A/r_B$ exceeds 1.46, the ordered pyrochlore structure is favored, with larger cations occupying the A sublattice and smaller cations occupying the B sublattice [9,10].

The literature generally agrees that the averaged cation-radius ratio $\overline{r_A}/\overline{r_B} \approx 1.46$ criterion remains broadly effective for distinguishing fluorite and pyrochlore phases in high-entropy CCCs [11,12]. Another descriptor, the modified size disorder $\delta_P^*$, originally introduced to predict reduced thermal conductivity in high-entropy pyrochlore CCCs, has also been introduced to forecast the phase formation in high-entropy fluorite-based CCCs [13]. The parameter $\delta_P^*$ was modified to account for the two distinct cation sublattices in the pyrochlore structure and is denoted by the subscript "*P*" to distinguish it from the size disorder parameter for the fluorite structure $\delta_F$, which contains only a single cation sublattice and will be discussed later. Moreover, an additional fluorite-pyrochlore dual-phase region was observed when $1.40 < \overline{r_A}/\overline{r_B} < 1.50, \delta_P^* \geq \sim 5\ \%$ [11,12,14]. However, several questions remain. First, previous studies using the modified size disorder $\delta_P^*$ implicitly assume a pyrochlore structure, thereby providing only a partial view of phase stability. In contrast, the size-disorder parameter $\delta_F$ provides a more physically reasonable descriptor for predicting phase stability, because it does not presuppose cation sublattice splitting. Instead, it captures the onset of sublattice splitting when the degree of cation disorder or size mismatch exceeds a critical threshold. Second, it remains unclear whether these descriptors are still valid as the compositional complexity increases. Some studies have found that phase stability rules based on the averaged cation-radius ratio are already shifted in UECs and non-equimolar systems [15–17]. Moreover, UECs have been reported to exhibit a surprisingly strong capability for single-phase formation in some cases [17]. Whether this behavior is broadly applicable across other UEC systems, however, remains to be established.

Extensive research has explored the use of the pyrochlore–fluorite ODTs to tune thermal conductivities in CCCs, covering systems from simple ternaries to complex UECs. The notable drop in thermal conductivity observed during ODTs has been attributed to the formation of short-range ordered nanodomains [15,16]. These studies have typically designed compositional series in either stoichiometric ($3+:4+ = 2:2$) or non-stoichiometric ($3+:4+ \neq 2:2$). Recently, one study systematically investigated the ODT in a non-stoichiometric fluorite–pyrochlore–weberite series, 20CCFBO$_{x\text{Nb/Ta}}$ $(15RE_{1/15})_{2x+1}(Ce_{1/3}Zr_{1/3}Hf_{1/3})_{3-3x}(Nb_{1/2}Ta_{1/2})_x O_{8-\delta}$ series ($0 \leq x \leq 1$). Linear projection of XRD relative peak intensities *vs.* compositional variables *x* was employed to predict the critical point of phase transitions, revealing abrupt fluorite–pyrochlore and pyrochlore-weberite transitions accompanied by discontinuous changes in structural order parameters [17]. Despite these advances, a fundamental question remains unaddressed: How does non-stoichiometry influence the nature of the ODT in CCCs?

In this study, we systematically investigated phase stability and ODTs in ultrahigh-entropy CCCs by synthesizing and characterizing three groups of 32 compositions containing 16-19 constituent cation components. The first group comprises 15 ultrahigh-entropy CCCs, $(15\mathrm{RE}_{1/15})_2\mathrm{TM}_2\mathrm{O}_7$, $(15\mathrm{RE}_{1/15}$ = equimolar mixture of 15 rare-earth elements; TM = Ce, Zr, Hf, Ti, or their equimolar combinations), which predominantly form single fluorite and pyrochlore phases, along with two fluorite–fluorite dual phases observed here for the first time. This group exhibits an unexpected propensity for cation ordering compared with conventional ternary and five-component high-entropy $A_2B_2O_7$ systems reported in the literature, indicating a fundamentally altered phase stability landscape.

Two additional 19-component series further clarify the role of stoichiometry in governing pyrochlore–fluorite ODTs: a stoichiometric series, $(15\mathrm{RE}_{1/15})_2[(\mathrm{Ce}_{1/3}\mathrm{Zr}_{1/3}\mathrm{Hf}_{1/3})_y\mathrm{Ti}_{1-y}]_2\mathrm{O}_7$ ($0 \leq y \leq 1$; denoted as "19CCC-P|F$_y$"), and a non-stoichiometric series, $(15\mathrm{RE}_{1/15})_{2-z}(\mathrm{CeZrHf})_{(1+z)/2}\mathrm{Ti}_{(1-z)/2}\mathrm{O}_{8-\delta}$ ($0 \leq z \leq 1$; denoted as "19CCC-P|F$_z$"). While both series' critical points were predicted using linear projections, the stoichiometric series conforms to the prediction, exhibiting a continuous ODT, whereas the non-stoichiometric series deviates markedly, displaying an abrupt transition.

Together, these results establish a systematic understanding of phase stability and ODTs behavior in ultrahigh-entropy CCCs and highlight stoichiometry as a critical factor.

## 2. Experimental procedure

### 1.1 Composition design

We designed and fabricated 15 ultrahigh-entropy CCCs containing 16-19 components, $(15\mathrm{RE}_{1/15})_2\mathrm{TM}_2\mathrm{O}_7$ (15RE = $La_{1/15}Pr_{1/15}Nd_{1/15}Sm_{1/15}Eu_{1/15}Gd_{1/15}Tb_{1/15}Dy_{1/15}Y_{1/15}Ho_{1/15}Er_{1/15}Tm_{1/15}Yb_{1/15}Lu_{1/15}Sc_{1/15}$; TM = Ce, Zr, Hf, Ti, or their equimolar combinations). In addition, we designed and fabricated two new 19-component series with nominal compositions:

1. Stoichiometric series "19CCC-P|F$_y$": $(15\mathrm{RE}_{1/15})_2[(\mathrm{Ce}_{1/3}\mathrm{Zr}_{1/3}\mathrm{Hf}_{1/3})_y\mathrm{Ti}_{1-y}]_2\mathrm{O}_7$ ($0 \leq y \leq 1$).

2. Non-stoichiometric series "19CCC-P|F$_z$": $(15\mathrm{RE}_{1/15})_{2-z}(\mathrm{CeZrHf})_{(1+z)/2}\mathrm{Ti}_{(1-z)/2}\mathrm{O}_{8-\delta}$ $(0 \leq z \leq 1)$.

These series were derived by mixing the single-phase pyrochlore endmembers $(15RE_{1/15})_2Ti_2O_7$, $(15RE_{1/15})_2(Ce_{1/4}Zr_{1/4}Hf_{1/4}Ti_{1/4})_2O_7$ with, a *y*-fraction of the single-phase fluorite endmembers $(15RE_{1/15})_2(Ce_{1/3}Zr_{1/3}Hf_{1/3})_2O_7$, and a *z*-fraction of single-phase fluorite endmembers $(15RE_{1/15}CeZrHf)O_{7.5}$, respectively.

The independent compositional variables *y* and *z* are introduced to distinguish them from the previously defined compositional variable *x* in a non-stoichiometric 20CCFBO$_{x\mathrm{Nb/Ta}}$ series $(15RE_{1/15})_{2x+1}(Ce_{1/3}Zr_{1/3}Hf_{1/3})_{3-3x}(Nb_{1/2}Ta_{1/2})_xO_{8-\delta}$, where *x* denotes the fraction of $Nb_{1/2}Ta_{1/2}$, and to facilitate direct comparison and discussion between the studies [17].

The compositions were tuned as follows:

- For the stoichiometric series "19CCC-P|F$_y$", $y = 0$, 0.375, 0.5625, 0.75, 0.8125, 0.8281, 0.8438, 0.875, 1.
- For the non-stoichiometric series "19CCC-P|F$_z$", $z = 0$, 0.125, 0.3125, 0.375, 0.4063, 0.4375, 0.4688, 0.5, 0.75, 1, where $y = 0.75$ corresponds to $z = 0$.

**2.1 Materials and synthesis**

We first prepared three mixtures of $(15RE_{1/15})_2O_3$, $(\mathrm{Ce}_{1/3}\mathrm{Zr}_{1/3}\mathrm{Hf}_{1/3})O_2$, and $(\mathrm{Ce}_{1/4}\mathrm{Zr}_{1/4}\mathrm{Hf}_{1/4}Ti_{1/4})O_2$ using commercially available high-purity binary oxide powders ($La_2O_3$, $Pr_6O_{11}$, $Nd_2O_3$, $Sm_2O_3$, $Eu_2O_3$, $Gd_2O_3$, $Tb_4O_7$, $Dy_2O_3$, $Y_2O_3$, $Ho_2O_3$, $Er_2O_3$, $Tm_2O_3$, $Yb_2O_3$, $Lu_2O_3$, $Sc_2O_3$, $CeO_2$, $ZrO_2$, $HfO_2$, and $TiO_2$; US Research Nanomaterials, TX, USA) with a particle size of ~ 5 µm and a purity > 99%. For each mixture, 10 g of powders were weighed in stoichiometric proportions and loaded into a 100 mL YSZ ball mill jar with a ball-to-powder ratio of 10:1. An appropriate amount of isopropyl alcohol (IPA) was added to form slurries. The powders were homogenized in a planetary ball mill (PQN04 gear-drive mill, Across International LLC, USA) at a rotation speed of 300 rpm for 24 h, followed by drying in an oven at 75 °C overnight.

To fabricate 15 ultrahigh-entropy CCCs $(15\mathrm{RE}_{1/15})_2\mathrm{TM}_2\mathrm{O}_7$ and two new series CCCs 19CCC-P|F$_y$ and 19CCC-P|F$_z$, the milled precursor mixtures were either weighed with $CeO_2$, $ZrO_2$, $HfO_2$, and $TiO_2$ powders, or combined in 2 g batches according to the designed stoichiometry. Each batch were load into a poly (methyl methacrylate) high-energy ball mill (HEBM) vial with two tungsten carbide (WC) inserts and a Ø 5/16" (~ 7.94 mm) WC ball. Approximately 1 wt % stearic acid was added as a process control agent. The vials were then placed in a SPEX 8000D mill (SPEXCertPrep, NJ, USA) and HEBM for 100 min.

The as-milled powders were uniaxially pressed into green pellets under ~ 100 MPa using a hydraulic press. The green pellets were placed on a Pt foil in an alumina crucible and sintered in a high-temperature box furnace (SentroTech, OH, USA) at 1600 °C for 24 h with a ramping rate of 5 °C/min, followed by furnace cooling. The sintered pellets were subsequently grinded and polished for further characterization.

**2.2 Characterization**

X-ray diffraction (XRD) patterns were acquired using a Rigaku GIXRD diffractometer operating at 40 kV and 44 mA, in a continuous scan mode with Cu Kα radiation. The full XRD patterns were obtained with a step size of 0.02 ° and a scan speed of 2.3 °/min. Slow scans were performed to resolve characteristic peaks at specific $2\theta$ degree ranges in a step scan mode, using a step size of 0.005 ° and a scan speed of 0.2 °/min. For compositions near the transition points, an even slower scan speed of 0.1 °/min was employed. The lattice parameters were determined via unit cell refinements using JADE software.

A Raman microscope (Renishaw inVia, England, UK) was used to further characterize ODTs from the selected polished specimens of the 19CCC-P|$F_y$ and 19CCC-P|$F_z$ series. A 633 nm He–Ne laser was employed with an estimated focused spot size of ~ 2 µm using a 20× objective.

Microstructural and compositional analyses were performed using scanning electron microscopy (SEM) and energy-dispersive X-ray spectroscopy (EDS) on a ThermoFisher (formerly FEI) Scios Dual Beam FIB/SEM equipped with an EDX detector at 20 kV.

## 3. Results and discussion

**3.1 Formation of ultrahigh-entropy CCCs**

The designed compositions and phases; the averaged cation-radius ratio $\overline{r_A}/\overline{r_B}$; size-disorder parameters ($\delta_F$ and $\delta_P^*$, with $\delta_P^*/\sqrt{2}$ also listed for comparison with $\delta_F$); measured and relative densities; and the ideal configurational entropies of mixing per cation ($\Delta S_{conf.}^{ideal}$) for the first group of 15 ultrahigh-entropy CCCs with nominal formula $(15RE_{1/15})_2TM_2O_7$ were characterized, calculated, and summarized in Table 1. Phase formation was determined by XRD (Figure 1), revealing five single-phase fluorite compositions, seven single-phase pyrochlore compositions, and three dual- or multiphase compositions. Two fluorite–fluorite dual-phase compositions, $(15RE_{1/15})_2Hf_2O_7$ and $(15RE_{1/15})_2(Zr_{1/2}Hf_{1/2})_2O_7$ (denoted as "F+F"), were further examined using slow step-scan XRD in the $2\theta$ range of 32–38°. The resulting patterns (Figure 1(d)) clearly reveal the coexistence of two distinct fluorite $(200)_F$ peaks, with no detectable reflection near $2\theta \approx 37°$, where the characteristic pyrochlore $(331)_P$ superstructure peak would normally appear. To our knowledge, this is the first report of fluorite-fluorite high-entropy dual phases.

Figure S1 shows representative cross-sectional SEM backscattered electron (BSE) images of multiphase $(15RE_{1/15})_2(Ce_{1/2}Ti_{1/2})_2O_7$ specimen and the fluorite-fluorite dual-phase $(15RE_{1/15})_2Hf_2O_7$ and $(15RE_{1/15})_2(Zr_{1/2}Hf_{1/2})_2O_7$ specimens. Distinct dark and bright contrasts are observed in the BSE images, where the dark regions correspond to phases enriched in lighter (lower-Z) elements, while the bright regions correspond to phases enriched in heavier (higher-Z) elements. The bright and dark regions are uniformly distributed throughout the specimens.

The EDS elemental mappings (Figure S2-S4) were conducted on the three dual-phase and multiphase specimens to examine elemental distributions within each phase. For the fluorite-fluorite dual-phase specimens $(15RE_{1/15})_2Hf_2O_7$ and $(15RE_{1/15})_2(Zr_{1/2}Hf_{1/2})_2O_7$, the EDS elemental maps (Figure S2(a) and S3(a)) indicates that each fluorite phase is compositionally homogeneous. No significant compositional differences between the two fluorite phases are detected within the noise level of EDS measurements. This conclusion is further supported by EDS point analysis acquired from the dark and bright regions (Figure S2(b) and S3(b)), which shows that the average compositions of the two fluorite phases are essentially identical within measurement errors.

For the multiphase specimen $(15RE_{1/15})_2(Ce_{1/2}Ti_{1/2})_2O_7$ (Figure S4), the EDS elemental maps reveal clear segregation, with Ce, Sm, Nd, and Ti enriched in regions of distinct contrast. However, unambiguous identification of the three constituent phases (fluorite, pyrochlore, and rhombohedral) in this composition based solely on SEM BSE images and EDS elemental maps is challenging and is not the focus of this work.

Figure 2 (a, b) presents plots of the averaged cation-radius ratio $\overline{r_A}/\overline{r_B}$ *vs.* the modified size disorder $\delta_P^*$, assuming an $A_2B_2O_7$ pyrochlore structure, and *vs.* size disorder $\delta_F$, assuming a disordered fluorite structure, for two groups of stoichiometric ultrahigh-entropy CCCs (A:B = 2:2): (1) 15 ultrahigh-entropy $(15RE_{1/15})_2TM_2O_7$ CCCs, and (2) the 19CCC-P|F$_y$ series,

$(15RE_{1/15})_2[(Ce_{1/3}Zr_{1/3}Hf_{1/3})_yTi_{1-y}]_2O_7$ $(0 \leq y \leq 1)$, which is further discussed in Section 3.2.

The modified size disorder $\delta_P^*$ follows a previously established definition based on the $A_2B_2O_7$ pyrochlore structure, where larger 3+ rare-earth cations occupy the eightfold-coordinated A sites and smaller 4+ cations occupy the sixfold-coordinated B sites, in the absence of anti-site defects [18].

To provide a more comprehensive evaluation of phase stability, a new parameter, size disorder $\delta_F$, is introduced:

$$\delta_F = \sqrt{\sum_{i=0}^{n} x_i(1 - r_i/\bar{r})^2}$$

This parameter assumes a disordered fluorite structure in which all cations are randomly mixed on a single lattice, and both 3+ and 4+ cations adopt their eightfold coordination radii.

As shown in Figure 2 (a,b) blue circular symbols represent single-phase fluorites; half–dark-blue and half–light-blue circular symbols represent dual-phase fluorites; red square symbols represent single-phase pyrochlores, and star-shaped symbols represent multiphase composition. In Figure 2(a), the typical fluorite–pyrochlore transition threshold ($r_A/r_B = 1.46$) for ternary $A_2B_2O_7$ systems are indicated by the black lines. The gray-shaded region marks the previously proposed fluorite–pyrochlore dual-phase stability window for high-entropy $A_2B_2O_7$ compositions ($1.40 \leq \overline{r_A}/\overline{r_B} \leq 1.50$, $\delta_P^* > 5$) [11,12,14].

As shown in Figure 2(a), at smaller value of $\delta_P^*$, the fluorite phase remain stable until $\overline{r_A}/\overline{r_B} \approx 1.46$, as evidenced by the fluorite-fluorite dual phase specimens $(15RE_{1/15})_2Hf_2O_7$, and $(15RE_{1/15})_2(Zr_{1/2}Hf_{1/2})_2O_7$. However, at larger $\delta_P^*$ values ($\delta_P^* \approx 13.87$), an ultrahigh-entropy pyrochlore phase emerges at $\overline{r_A}/\overline{r_B} \approx 1.40$ (corresponding to the specimen $y$ = 0.8281 in the 19CCC-P|F$_y$ series). These observations indicate a stronger ordering tendency in ultrahigh-entropy CCCs than in ternary and high-entropy CCCs.

Shifted ODT thresholds have been reported across different ultrahigh-entropy fluorite-based CCCs series: an 12-component CCCs series ($[(Sm_{0.25}Eu_{0.25}Gd_{0.25}Yb_{0.25})_2(Ti_{0.5}Hf_{0.25}Zr_{0.25})_2O_7]_{1-x}[(Sc_{0.266}Dy_{0.248}Tm_{0.246}Yb_{0.240})_3NbO_7]_x, 0 \leq x \leq 1$) exhibits an ODT at a higher threshold of $\overline{r_A}/\overline{r_B} \approx 1.48$, while a 10-component CCCs series ($[(Pr_{0.375}Nd_{0.375}Yb_{0.25})_2(Ti_{0.5}Hf_{0.25}Zr_{0.25})_2O_7]_{1-x}[(DyHoErNb)O_7]_x, 0 \leq x \leq 1$) undergoes

ODT at a much lower threshold of $\overline{r_A}/\overline{r_B} \approx 1.30$. More recently, a 20-component CCC series ($(15RE_{1/15})_{2x+1}(Ce_{1/3}Zr_{1/3}Hf_{1/3})_{3-3x}(Nb_{1/2}Ta_{1/2})_xO_{8-\delta}, 0 \le x \le 1$) was found to undergo ODT at $\overline{r_A}/\overline{r_B} \approx 1.41$. These observations collectively suggest that the averaged cation-radius ratio $\overline{r_A}/\overline{r_B}$ is no longer a fully reliable phase-stability criterion in ultrahigh-entropy systems .

When $\delta_P^*$ increases further, a multi-phase (fluorite, pyrochlore, and $R\bar{3}m$ rhombohedral phase) forms in the $(15RE_{1/15})_2(Ce_{1/2}Ti_{1/2})_2O_7$ specimen at $\overline{r_A}/\overline{r_B} \approx 1.40$ and $\delta_P^* \approx 19.13$. This is not unexpected, as greater size mismatches correspond to increased lattice strain, which can promote phase separation.

However, somewhat surprisingly, although the $\delta_P^*$ value in these ultrahigh-entropy CCCs are generally larger than those reported for high-entropy CCCs [11,12,14], the formation of fluorite-pyrochlore dual phases is noticeably suppressed within the conventional dual-phase stability window ($1.40 \le r_A/r_B \le 1.50, \delta_P^* > 5$).

When the averaged cation-radius ratio $\overline{r_A}/\overline{r_B}$ is plotted against the size disorder $\delta_F$ (Figure 2(b)), it is evident that $\delta_F$ provides a more direct and reliable criterion, than either $\overline{r_A}/\overline{r_B}$ or $\delta_P^*$, for defining the fluorite-pyrochlore transition threshold in ultrahigh-entropy CCCs. The fluorite-to-pyrochlore transition occurs at $\delta_F \approx 12.10$ (indicated by the vertical dashed line), with fluorite phases remaining stable below this value and pyrochlore or multiphase phases stabilizing above it.

### 3.2 Pyrochlore-fluorite ODT

#### 3.2.1 Pyrochlore-fluorite ODT in stoichiometric *vs.* non-stoichiometric CCCs series

Two new 19-component CCCs series, 19CCC-P|F$_y$ ($(15RE_{1/15})_2[(Ce_{1/3}Zr_{1/3}Hf_{1/3})_yTi_{1-y}]_2O_7$ , $0 \le y \le 1$) and 19CCC-P|F$_z$ ($(15RE_{1/15})_{2-z}(CeZrHf)_{(1+z)/2}Ti_{(1-z)/2}O_{8-\delta}$, $0 \le z \le 1$), were designed and synthesized to investigate the influence of stoichiometry (maintaining or deviating from an ideal 2:2 A/B cation ratio) on phase stability and ODTs in ultrahigh-entropy CCCs. Figure 3 schematically illustrates the compositions investigated in both series. Table 2 and Table 3 summarize the designed compositions, phase, the averaged cation-radius ratio $\overline{r_A}/\overline{r_B}$, size-disorder parameters ($\delta_F$, $\delta_P^*$, and $\delta_P^*/\sqrt{2}$), lattice parameter, measured densities, relative densities, and ideal configurational entropies of mixing per cation ($\Delta S_{conf.}^{ideal}$) for all compositions in the two new CCCs series.

Full XRD patterns confirm that all compositions in both series form exclusively single-phase fluorite or pyrochlore structures (Supplementary Figures S5-S6), showing a strong capability for single-phase formation similar to what we observed in other ultrahigh-entropy systems [17,19]. To examine the pyrochlore-fluorite ODTs in greater detail, slow XRD scans were performed over $2\theta \approx 32-38°$, encompassing the fluorite $(200)_F$, the pyrochlore $(400)_P$, and the characteristic pyrochlore $(331)_P$ superstructure peaks (Figure 4(a) and Figure 5(a)). Enlarged views of the pyrochlore $(331)_P$ superstructure peaks in Figure 4(b) and Figure 5(b) clearly shows that the $(331)_P$ reflection progressively weakens and eventually disappears with increasing compositional variables $y$ and $z$ (*i.e.* increasing $y$- and $z$- fraction of fluorite endmembers). The pyrochlore-to-fluorite ODTs are identified at $y \approx 0.8125{\sim}0.8281$ for 19CCC-P|$F_y$ series and $z \approx 0.4063{\sim}0.4375$ for 19CCC-P|$F_z$ series.

Figure 6 presents the full width at half maximum (FWHM) analysis for the fluorite $(200)_F$, pyrochlore $(440)_P$, and characteristic pyrochlore $(331)_P$ peaks for the stoichiometric 19CCC-P|$F_y$ (Figure 6(a, c)) and non-stoichiometric 19CCC-P|$F_z$ series (Figure 6(b, d)). The results show that, as the compositions approach the ODT, the pyrochlore $(331)_P$ peaks undergoes a sudden broadening, while the pyrochlore $(400)_P$ peaks initially narrow and then gradually broaden as structural disorder increase with the compositional variable $y$ and $z$, respectively. These observations suggest that, during the ODT, $(331)_P$ planes experience increased lattice strain, while the $(400)_P$ planes undergo a reduction in lattice strain. Overall, FWHM evolution provides complementary information on the phase evolution during the ODT process beyond that observed from XRD.

Linear projections were performed using the pyrochlore superstructure peak intensity $(331)_P/(440)_P$ as an order parameter to quantitatively estimate critical points of these ODTs (Figure 4(c) and 5(c)). Six data points from each series were used: $y$ = 0, 0.375, 0.5625, 0.75, 0.8125, and 0.8281 for the stoichiometric 19CCC-P|$F_y$ series, and $z$ = 0, 0.125, 0.25, 0.3125, 0.375, and 0.4063 for the non-stoichiometric 19CCC-P|$F_z$ series. The linear regression yielded the following relationships:

Stoichiometric 19CCC-P|$F_y$ series: $y = -0.5954x + 0.4978$ ($R^2 = 0.9937$)

Non-stoichiometric 19CCC-P|$F_z$ series: $y = -0.0664x + 0.0424$ ($R^2 = 0.9683$)

These results suggest a projected pyrochlore-fluorite ODT at $y \approx 0.84$ for the stoichiometric 19CCC-P|$F_y$ series, in good agreement with the experimentally observed ODT at $y = 0.8438$. These demonstrate an excellent agreement between the predicted and measured transition compositions in the stoichiometric series.

For the non-stoichiometric 19CCC-P|$F_z$ series, the linear projection predicts an ODT at $z \approx 0.64$. However, the experimentally observed ODT occurs at $z \approx 0.41 \sim 0.44$, indicating a significantly more abrupt transition than projection, as evidenced by the rapid disappearance of the characteristic pyrochlore $(331)_P$ superstructure peaks near ODT critical points (Figure 5(b); z = 0.4063 and z = 0.4375). A similarly abrupt ODT behavior has been reported in another non-stoichiometric 20CCFBO$_{x\mathrm{Nb/Ta}}$ series ($(15RE_{1/15})_{2x+1}(Ce_{1/3}Zr_{1/3}Hf_{1/3})_{3-3x}(Nb_{1/2}Ta_{1/2})_x O_{8-\delta}$ , $0 \leq x \leq 1$) , where the linear regression of pyrochlore superstructure peak intensity *vs.* compositional variable $x$ yielded: $y = -0.0089 + 0.0623x, R^2 = 0.9875$, suggesting a projected pyrochlore-fluorite ODT at $x \approx 0.17$. This projection is also considerably lower than the experimentally observed ODT at $x \approx 0.27$. Together, these results suggest an abrupt nature of the pyrochlore–fluorite ODTs in non-stoichiometric CCCs series.

Raman spectroscopy is widely recognized as an effective probe for distinguishing ordered, characterized by well-resolved vibrational bands, from disordered structures, which exhibit broadened spectra [20]. It therefore provides structural information complementary to XRD for investigating the ODTs in the 19CCC-P|$F_y$ and 19CCC-P|$F_z$ series. Figure 7 presents selected Raman spectra from both series, including the fluorite and pyrochlore endmembers ($y = 0, 1$ and $z = 0, 1$), as well as compositions immediately before and after the pyrochlore-to-fluorite ODTs ($y = 0.8281, 0.8438$ and $z = 0.4063, 0.4375$).

As shown in Figure 7(a) for the 19CCC-P|$F_y$ series, the ultrahigh-entropy pyrochlore titanate endmember ($y = 0$) exhibits prominent $F_{2g}$, $E_g$, and $A_{1g}$ modes, confirming that this composition adopts a well-ordered pyrochlore structure. As the composition approaches the pyrochlore-to-fluorite ODT critical point, the Raman bands for $y = 0.8281$ progressively broaden but still retain features characteristic of the pyrochlore structure. Once the ODT is completed, $y = 0.8438$ composition exhibits a loss of spectral resolution, indicating the breakdown of long-range order. The fluorite endmember ($y = 1$) likewise exhibits a broadened spectral profile.

In contrast, for the 19CCC-P|$F_z$ series (Figure 7(b)), the ultrahigh-entropy pyrochlore endmember ($z = 0$) exhibits only weak features reminiscent of the pyrochlore structure, rather than well-resolved characteristic modes. This behavior is expected given the non-stoichiometric nature of the composition, which suppresses the formation of a well-crystallized pyrochlore

structure. A similar trend of spectral broadening is observed with increasing $z$, indicating progressive structural disorder.

Overall, the evolution from well-resolved spectra to broadened features across the pyrochlore-to-fluorite ODTs is consistent with the disordering behavior commonly reported for ternary pyrochlores in the literature. However, the Raman spectra does not provide additional quantitative information on the ODTs beyond that obtained from XRD.

**3.2.2 Phase stability and descriptors**

The phase stability was further evaluated through a combined analysis of multiple variables: the compositional variables $y$ and $z$, the averaged cation-radius ratio $\overline{r_A}/\overline{r_B}$, the size-disorder parameters $\delta_F$ and $\delta_P^*$, and the non-stoichiometric parameter $t$ (Figure 8). The non-stoichiometric parameter $t$ describes compositions of the form $A_{2-2t}B_{2+2t}O_7$, quantifying the deviation from the ideal A/B = 2:2 cation ratio. For the stoichiometric 19CCC-P|F$_y$ series, $(15\mathrm{RE}_{1/15})_2[(\mathrm{Ce}_{1/3}\mathrm{Zr}_{1/3}\mathrm{Hf}_{1/3})_y\mathrm{Ti}_{1-y}]_2\mathrm{O}_7$, $t = 0$, while for the non-stoichiometric series 19CCC-P|F$_z$ series, $(15\mathrm{RE}_{1/15})_{2-z}(\mathrm{CeZrHf})_{(1+z)/2}\mathrm{Ti}_{(1-z)/2}\mathrm{O}_{8-\delta}$, $t = z/2$.

As shown in Figure 8(a, b), the pyrochlore-fluorite ODTs occur at averaged cation-radius ratios of $\overline{r_A}/\overline{r_B} = 1.40$ for the stoichiometric 19CCC-P|F$_y$ and $\overline{r_A}/\overline{r_B} = 1.42$ for the non-stoichiometric 19CCC-P|F$_z$ series (see black dashed lines; calculated by forcing an A/B = 2:2 ratio). Both values are notably lower than the established threshold for ternary zirconates and hafnates ($r_A/r_B = 1.46$, black solid line). Furthermore, the linear-projection-predicted critical composition for the pyrochlore-to-fluorite transition in 19CCC-P|F$_z$ series occurs at $z = 0.64$, corresponding to $\overline{r_A}/\overline{r_B} = 1.41$, which is also significantly below the conventional threshold of $r_A/r_B = 1.46$. These indicate that ultrahigh-entropy CCCs exhibit a stronger tendency toward ordering compared to ternary ceramics. Comparison of the pyrochlore-fluorite ODT thresholds in terms of $\overline{r_A}/\overline{r_B}$ for the stoichiometric 19CCC-P|F$_y$ series and non-stoichiometric 19CCC-P|F$_z$ series indicates that the stoichiometric series is more prone to ordering. This is consistent with expectations, as an $A/B = 2:2$ stoichiometric ratio promotes the preferential occupation of 3+ and 4+ cations on the A and B sites, respectively, in the typical $A_2^{3+}B_2^{4+}O_7$ pyrochlore structure.

The averaged cation-radius ratio $\overline{r_A}/\overline{r_B}$ *vs.* size disorder $\delta_F$ and $\delta_P^*$ are shown in Figure 8(c) and (d), respectively. Both sets of descriptors clearly define that the fluorite phases are stable near the lower-left corner, corresponding to small $\overline{r_A}/\overline{r_B}$ (favor cations mixing in a single lattice) coincide with small $\delta_F$, and $\delta_P^*$ (corresponding to lower lattice strain and enhanced lattice mixing), while the pyrochlore phases are stable near the upper-right corner with larger values of these parameters. The dashed lines, which connect the ODT critical points in both series, delineate the boundary between the regions where the fluorite phase is stable and those where the pyrochlore phase is stable.

Figure 8(e) shows the modified size disorder $\delta_P^*$ *vs.* size disorder $\delta_F$ diagram. It indicates that the fluorite phases are stable in the region with smaller $\delta_F$, whereas the pyrochlore phase is table in the region with larger $\delta_F$, with no pronounced trend associated with increasing $\delta_P^*$. This observation suggests that $\delta_F$ is the more effective descriptor for distinguishing the two phases. The dashed line delineates the phase stability boundary between these stability regions.

Figure 8(f) presents a diagram of the averaged cation-radius ratio $\overline{r_A}/\overline{r_B}$ *vs.* non-stoichiometry parameter $t$. The parameter $t$ quantifies the degree of deviation from the ideal 2:2 A/B ratio and is defined for both series by the general composition $A_{2-2t}B_{2+2t}O_{8-\delta}$. For the stoichiometric 19CCC-P|F$_y$ series, $t = 0$, whereas for the non-stoichiometric 19CCC-P|F$_z$ series, $t = z/2$. As shown in the figure, it is evident that $\overline{r_A}/\overline{r_B}$ plays a more dominant role than $t$ in determining phase stability.

In the non-stoichiometric 19CCC-P|F$_z$ series, the pyrochlore phases remains stable up to $t = 0.20$, corresponding to an ODT critical A/B ratio of 0.66 (A-site deficient). Beyond this threshold, progressive disordering occurs. The influence of non-stoichiometry on the ODT has been discussed in different CCCs systems. In ternary rare earth zirconates and hafnates $A_{2\pm t}B_{2\mp t}O_{7\mp t/2}$, the ODT occurs when $|t| > 0.1 - 0.2$, corresponding to a critical A/B ratio of $0.82 - 1.22$[21]. In a 11-component CCCs series $[(Sm_{0.25}Eu_{0.25}Gd_{0.25}Yb_{0.25})_2(Ti_{0.5}Hf_{0.25}Zr_{0.25})_2O_7]_{1-x}[(Sc_{0.266}Dy_{0.248}Tm_{0.246}Yb_{0.240})_3NbO_7]_x$, the ODT was reported at a critical A/B ratio between 1.16 and 1.20 [16]. In a recent study of a 20-component 20CCFBO$_{x\mathrm{Nb/Ta}}$ series $(15RE_{1/15})_{2x+1}(Ce_{1/3}Zr_{1/3}Hf_{1/3})_{3-3x}(Nb_{1/2}Ta_{1/2})_xO_{8-\delta}$, the pyrochlore structure remained stable until $A/B = 0.6$[17]. These observations suggest that ultrahigh-entropy CCCs are better able to tolerate non-stoichiometry while maintaining an ordered structure compared with ternary CCCs.

We computed and plotted ideal configurational entropy of mixing *vs.* compositional variable $y$ and $z$ for stoichiometric 19CCC-P|F$_y$ and non-stoichiometric 19CCC-P|F$_z$ series, respectively, as shown in Figure 9. For the pyrochlore phase with two sublattices, we calculated and plotted the mixing entropy for each sublattice separately ($\Delta S_{A\,site}^{mix}$ and $\Delta S_{B\,site}^{mix}$). We then plotted weighted average entropies for the $A_2B_2O_7$ pyrochlore phase using: $\Delta\bar{S}_P^{mix} = \frac{1}{2}\Delta S_{A\,site}^{mix} + \frac{1}{2}\Delta S_{B\,site}^{mix}$. The resulting weighted average ideal configurational entropy of mixing is generally greater than $2k_B$ per cation for both the stoichiometric 19CCC-P|F$_y$ and non-stoichiometric 19CCC-P|F$_z$ series, indicating that these compositions fall within the ultrahigh-entropy CCCs, even though short-range ordering (if present) may reduce the actual configurational entropies for fluorite composition. The weighted average configurational entropy $\Delta\bar{S}_P^{mix}$ reaches its maximum near ODT critical point in the stoichiometric 19CCC-P|F$_y$ series, suggesting that the ODT is possible entropy-driven (primarily driven from the B-site configurational entropy $\Delta S_{B\,site}^{mix}$). In contrast, $\Delta\bar{S}_P^{mix}$ shows no clear correlation with phase formation or the ODT in the non-stoichiometric 19CCC-P|F$_z$ series, consistent with our previous observations in another non-stoichiometric 20CCFBO$_{xNb/Ta}$ series [17]. It is reasonable, as oxygen vacancies and anti-site defect in the non-stoichiometric 19CCC-P|F$_z$ series complicate accurate estimation of the configurational entropy. In addition, the difference in configurational entropies of mixing during the ODT, $\Delta\bar{S}_F^{mix} - \Delta\bar{S}_P^{mix}$, was calculated and is indicated by black arrows in the figure. For the stoichiometric 19CCC-P|F$_y$, $\Delta\bar{S}_F^{mix} - \Delta\bar{S}_P^{mix} = 0.689\ k_B$ per cation, whereas for the non-stoichiometric 19CCC-P|F$_z$ series, $\Delta\bar{S}_F^{mix} - \Delta\bar{S}_P^{mix} = 0.555\ k_B$ per cation, indicating that the non-stoichiometric CCCs are more prone to disordering. The configurational entropy difference at the ODT in ternary CCCs is $0.693\ k_B$ pre cation. This value is close to what we observe in stoichiometric ultrahigh-entropy CCCs, suggesting an inherent tendency toward ordering in stoichiometric systems.

The lattice parameters, normalized to that of the primitive cubic fluorite cell ($a_f$) for stoichiometric 19CCC-P|F$_y$ and non-stoichiometric 19CCC-P|F$_z$ series are shown in Figure 10. In both series, the normalized lattice parameters generally increase gradually with the compositional variables, with a more pronounced trend observed in the stoichiometric 19CCC-P|F$_y$ series reflecting the larger lattice parameter difference between its endmembers.

## 4. Conclusion

In this work, we systematically synthesized and investigated three groups of 32 ultrahigh-entropy compositionally complex ceramics (CCCs) containing 16–19 cations. In the first group of 15 CCCs, $(15RE_{1/15})_2TM_2O_7$, we identified 12 single-phase ultrahigh-entropy fluorite or pyrochlore ceramics, along with two compositions exhibiting fluorite–fluorite dual phases—reported here for the first time. Compared with ternary $A_2B_2O_7$ (which undergo the fluorite–pyrochlore transition at $r_A/r_B \approx 1.46$), these ultrahigh-entropy CCCs show a stronger tendency toward ordering (pyrochlore formation), even at higher size disorder levels where disordering would typically be favored. Among the structural descriptors examined, $\delta_F$ is found to be more effective for predicting phase formation in ultrahigh-entropy CCCs than either $\overline{r_A}/\overline{r_B}$ or $\delta_P^*$. Moreover, no fluorite–pyrochlore dual-phase CCCs were detected within the conventional stability window ( $1.40 < \overline{r_A}/\overline{r_B} < 1.50, \delta_P^* > 5\ \%$ ) previously established for high-entropy CCCs, indicating a pronounced suppression of dual-phase formation in the ultrahigh-entropy regime. The pyrochlore-fluorite ODT was further investigated in two newly designed 16–19-component CCC series: the stoichiometric 19CCC-P|F$_y$ and the non-stoichiometric 19CCC-P|F$_z$ series, both of which form exclusively single-phase pyrochlore or fluorite structures across their entire compositional ranges. Phase evolution and transition behaviors were systematically characterized through detailed analysis of XRD peak features. In the stoichiometric 19CCC-P|F$_y$ series, the ODT occurs at a predictable composition consistent with the linear projection of the pyrochlore $(331)_P$ superstructure peak intensity used as an order parameter. In contrast, the non-stoichiometric 19CCC-P|F$_z$ series exhibits a significantly more abrupt transition. The effect of non-stoichiometry on phase stability and the ODT was evaluated through a combined analysis of multiple structural descriptors, including $\overline{r_A}/\overline{r_B}$, $\delta_F$, and $\delta_P^*$. Both the stoichiometric 19CCC-P|F$_y$ ($\overline{r_A}/\overline{r_B} \approx 1.40$) and the non-stoichiometric 19CCC-P|F$_z$ series ($\overline{r_A}/\overline{r_B} \approx 1.42$) exhibit a stronger propensity for ordering than ternary $A_2B_2O_7$ CCCs, which undergo the fluorite–pyrochlore transition at $r_A/r_B \approx 1.46$. This tendency is particularly pronounced in the stoichiometric 19CCC-P|F$_y$ series, which is consistent with the fact that half of the 3+ cations naturally occupy the A site and half of the 4+ cations occupy the B site in the ordered $A_2B_2O_7$pyrochlore structure.

**Acknowledgment:** The work is supported by the U.S. National Science Foundation (NSF), primarily through NSF Grant No. DMR-2026193 from July 2020 to March 2025, and we also thank partial support from NSF Future Manufacturing Research Grant No. CBET-2328044 after March 2025 to complete this study related to the general topic of ceramic processing and manufacturing.

**Table 1.** Designed compositions of the first group of 15 ultrahigh-entropy CCCs with the nominal formula $(15RE_{1/15})_2TM_2O_7$, where $15RE_{1/15}$ denotes an equimolar mixture of 15 rare-earth elements (Sc, Y, La, Pr, Nd, Sm, Eu, Gd, Tb, Dy, Ho, Er, Tm, Yb, and Lu), and TM = Ce, Zr, Hf, Ti, or their equimolar combinations. The table summarizes the observed phase, the averaged cation-radius ratio ($\bar{r}_A/\bar{r}_A$), and the size-disorder parameters for all 15 compositions. Here, "F" denotes a single-phase fluorite structure, "P" denotes a pyrochlore phase, "F+F" indicates dual-fluorite phases, and "F+P+$R\bar{3}m$" represents a three-phase mixture. Two size-disorder parameters were calculated for each composition (regardless the actual stable phase) for comparison: $\delta_F$ was calculated assuming a disordered fluorite structure, while $\delta_P^*$ was calculated assuming an ordered pyrochlore structure (following the definition of Wright *et al.* [18]). The ideal configurational entropies of mixing per cation ($\Delta S_{conf.}^{ideal}$) are also calculated for all single-phase specimens based on the actual stable phases observed.

| Composition | Phase | $\bar{r}_A/\bar{r}_B$ | Size Disorder | | | Measured Density (g/cm$^3$) | Relative Density (%) | $\Delta S_{conf.}^{ideal}$ ($k_B$/cation) | | | |
|---|---|---|---|---|---|---|---|---|---|---|---|
| | | | $\delta_F$ (%) | $\delta_P^*$ (%) | $\delta_P^*/\sqrt{2}$ (%) | | | Fluorite | Pyrochlore | | |
| | | | | | | | | | A-site | B-Site | Mean |
| $(15RE_{1/15})_2Ce_2O_7$ | F | 1.19 | 5.78 | 6.56 | 4.64 | 6.98 | 94.6 | 2.05 | | | |
| $(15RE_{1/15})_2Zr_2O_7$ | F | 1.44 | 11.59 | 6.56 | 4.64 | 6.75 | 96.7 | 2.05 | | | |
| $(15RE_{1/15})_2Hf_2O_7$ | F + F | 1.46 | 12.13 | 6.56 | 4.64 | 8.68 | - | | | | |
| $(15RE_{1/15})_2Ti_2O_7$ | P | 1.71 | 17.47 | 6.56 | 4.64 | 5.96 | 93.9 | | 2.71 | 0 | 1.35 |
| $(15RE_{1/15})_2(Ce_{1/2}Zr_{1/2})_2O_7$ | F | 1.30 | 9.58 | 11.49 | 8.12 | 6.81 | 94.9 | 2.05 | | | |
| $(15RE_{1/15})_2(Zr_{1/2}Hf_{1/2})_2O_7$ | F + F | 1.45 | 11.87 | 6.59 | 4.66 | 7.65 | - | | | | |
| $(15RE_{1/15})_2(Ce_{1/2}Hf_{1/2})_2O_7$ | F | 1.31 | 9.97 | 12.06 | 8.53 | 7.71 | 94.6 | 2.05 | | | |
| $(15RE_{1/15})_2(Ce_{1/2}Ti_{1/2})_2O_7$ | F + P + $R\bar{3}m$ | 1.40 | 13.80 | 19.13 | 13.52 | 6.68 | - | | | | |
| $(15RE_{1/15})_2(Zr_{1/2}Ti_{1/2})_2O_7$ | P | 1.56 | 14.93 | 10.88 | 7.69 | 6.45 | 98.2 | | 2.71 | 0.69 | 1.70 |
| $(15RE_{1/15})_2(Hf_{1/2}Ti_{1/2})_2O_7$ | P | 1.57 | 15.12 | 10.33 | 7.31 | 7.49 | 97.9 | | 2.71 | 0.69 | 1.70 |
| $(15RE_{1/15})_2(Ce_{1/3}Zr_{1/3}Hf_{1/3})_2O_7$ | F | 1.35 | 10.62 | 11.58 | 8.19 | 7.42 | 95.3 | 2.60 | | | |
| $(15RE_{1/15})_2(Ce_{1/3}Zr_{1/3}Ti_{1/3})_2O_7$ | P | 1.41 | 13.12 | 16.21 | 11.47 | 6.75 | 99.4 | | 2.71 | 1.10 | 1.90 |
| $(15RE_{1/15})_2(Ce_{1/3}Hf_{1/3}Ti_{1/3})_2O_7$ | P | 1.42 | 13.29 | 16.33 | 11.55 | 7.38 | 97.6 | | 2.71 | 1.10 | 1.90 |
| $(15RE_{1/15})_2(Zr_{1/3}Hf_{1/3}Ti_{1/3})_2O_7$ | P | 1.53 | 14.07 | 10.09 | 7.13 | 7.20 | 98.2 | | 2.71 | 1.10 | 1.90 |
| $(15RE_{1/15})_2(Ce_{1/4}Zr_{1/4}Hf_{1/4}Ti_{1/4})_2O_7$ | P | 1.43 | 12.89 | 14.56 | 10.30 | 7.18 | 98.2 | | 2.71 | 1.39 | 2.05 |

**Table 2.** Phases, averaged cation-radius ratios ($\bar{r}_A/\bar{r}_B$), two size-disorder parameters ($\delta_F$ and $\delta_P^*$), lattice parameters, measured and relative densities, and calculated ideal configurational entropies of mixing ($\Delta S_{conf.}^{ideal}$) for nine specimens in the second series of 19-component CCCs exhibiting ultrahigh-entropy fluorite or pyrochlore phases (denoted as "19CCC-P|F$_y$"), $[(15\text{RE}_{1/15})_2\text{Ti}_2\text{O}_7]_{1-y}[(15\text{RE}_{1/15})_2(\text{Ce}_{1/3}\text{Zr}_{1/3}\text{Hf}_{1/3})_2\text{O}_7]_y$, where $y$ ($0 \leq y \leq 1$) represents the fraction of the disordered defect-fluorite endmember, $(15\text{RE}_{1/15})_2(\text{Ce}_{1/3}\text{Zr}_{1/3}\text{Hf}_{1/3})_2\text{O}_7$. This chemical formula can also be expressed in simplified form as $(15\text{RE}_{1/15})_2[(\text{Ce}_{1/3}\text{Zr}_{1/3}\text{Hf}_{1/3})_y\text{Ti}_{1-y}]_2\text{O}_7$. For the ordered pyrochlore phase, $(1 - z)$ represents the fraction of the (smallest) $Ti^{4+}$ ions occupying the B site in the $A_2B_2O_7$ structure, and a lower Ti fraction (*i.e.*, larger *z*) decreases the $\bar{r}_A/\bar{r}_B$ ratio, thereby promoting disorder.

| $y$ | Phase | $\bar{r}_A/\bar{r}_A$ | Size Disorder | | | Lattice Parameter | Measured Density (g/cm$^3$) | Relative Density (%) | $\Delta S_{conf.}^{ideal}$ ($k_B$/cation) | | | |
|---|---|---|---|---|---|---|---|---|---|---|---|---|
| | | | $\delta_F$ (%) | $\delta_P^*$ (%) | $\delta_P^*/\sqrt{2}$ (%) | $a_F$ or $1/2a_P$ (Å) | | | Fluorite | Pyrochlore | | |
| | | | | | | | | | | A-site | B-Site | Mean |
| 0 | P | 1.711 | 17.47 | 6.56 | 4.64 | 5.0757 ± 0.0002 | 5.96 | 93.9 | | 2.71 | 0 | 1.35 |
| 0.375 | P | 1.555 | 15.50 | 15.05 | 10.64 | 5.1516 ± 0.0008 | 6.26 | 91.1 | | 2.71 | 1.07 | 1.89 |
| 0.5625 | P | 1.487 | 14.29 | 15.43 | 10.91 | 5.2062 ± 0.0025 | 7.04 | 99.9 | | 2.71 | 1.30 | 2.01 |
| 0.75 | P | 1.425 | 12.89 | 14.56 | 10.30 | 5.2196 ± 0.0008 | 7.18 | 98.2 | | 2.71 | 1.39 | 2.05 |
| 0.8125 | P | 1.405 | 12.37 | 14.02 | 9.92 | 5.2507 ± 0.0010 | 7.31 | 97.0 | | 2.71 | 1.38 | 2.04 |
| 0.8281 | P | 1.401 | 12.24 | 13.87 | 9.81 | 5.2227 ± 0.0077 | 7.20 | 95.6 | | 2.71 | 1.37 | 2.04 |
| 0.8438 | F | 1.396 | 12.10 | 13.71 | 9.69 | 5.2548 ± 0.0004 | 7.29 | 95.8 | 2.73 | | | |
| 0.875 | F | 1.386 | 11.83 | 13.36 | 9.44 | 5.2615 ± 0.0004 | 7.20 | 94.1 | 2.72 | | | |
| 1 | F | 1.350 | 10.62 | 11.58 | 8.19 | 5.2425 ± 0.0004 | 7.42 | 95.3 | 2.60 | | | |

**Table 3.** Phases, averaged cation-radius ratios ($\bar{r}_A/\bar{r}_B$), two size-disorder parameters ($\delta_F$ and $\delta_P^*$), lattice parameters, measured and relative densities, and calculated ideal configurational entropies of mixing ($\Delta S_{conf.}^{ideal}$) for 11 specimens in the third series of 19-component CCCs exhibiting ultrahigh-entropy fluorite or pyrochlore phases (denoted as "19CCC-P|F$_z$"), $[(15RE_{1/15})_2(Ce_{1/4}Zr_{1/4}Hf_{1/4}Ti_{1/4})_2O_7]_{1-z}[(15RE_{1/15}CeZrHf)O_{7.5}]_z$ , where $z$ ( $0 \leq z \leq 1$ ) represents the fraction of the disordered fluorite endmember, $(15RE_{1/15}CeZrHf)O_{7.5}$. This chemical formula can also be expressed in a simplified form as $(15RE_{1/15})_{2-z}(CeZrHf)_{(1+z)/2}Ti_{(1-z)/2}O_{8-\delta}$. For the ordered pyrochlore phase, $z/2$ also represents the fraction of the Ce (presumably $Ce^{4+}$) ions occupying the A site in the $A_2B_2O_7$ structure, and a larger value of $z$ promotes disorder.

| $z$ | Phase | $\bar{r}_A/\bar{r}_B$ | Size Disorder | | | Lattice Parameter | Measured Density (g/cm$^3$) | Relative Density (%) | $\Delta S_{conf.}^{ideal}$ ($k_B$/cation) | | | |
|---|---|---|---|---|---|---|---|---|---|---|---|---|
| | | | | | | | | | Fluorite | Pyrochlore | | |
| | | | $\delta_F$ (%) | $\delta_P^*$ (%) | $\delta_P^*/\sqrt{2}$ (%) | $a_F$ or 1/2 $a_P$ (Å) | | | | A-site | B-Site | Mean |
| 0 | P | 1.425 | 12.89 | 14.56 | 10.30 | 5.2196 ± 0.0008 | 7.18 | 98.2 | | 2.71 | 1.39 | 2.05 |
| 0.125 | P | 1.422 | 12.62 | 13.85 | 9.79 | 5.2374 ± 0.0003 | 7.38 | 100.0 | | 2.77 | 1.38 | 2.08 |
| 0.25 | P | 1.419 | 12.33 | 13.07 | 9.25 | 5.2412 ± 0.0011 | 7.28 | 98.4 | | 2.75 | 1.35 | 2.05 |
| 0.3125 | P | 1.418 | 12.17 | 12.66 | 8.95 | 5.2399 ± 0.0005 | 7.40 | 99.7 | | 2.72 | 1.34 | 2.03 |
| 0.375 | P | 1.417 | 12.02 | 12.23 | 8.65 | 5.2412 ± 0.0009 | 6.96 | 93.5 | | 2.68 | 1.31 | 2.00 |
| 0.4062 | P | 1.416 | 11.94 | 12.01 | 8.49 | 5.2404 ± 0.0013 | 7.29 | 97.6 | | 2.66 | 1.30 | 1.98 |
| 0.4375 | F | 1.415 | 11.86 | 11.78 | 8.33 | 5.2377 ± 0.0003 | 7.27 | 94.7 | 2.54 | | | |
| 0.4688 | F | 1.414 | 11.77 | 11.54 | 8.16 | 5.2395 ± 0.0002 | 7.41 | 96.5 | 2.52 | | | |
| 0.5 | F | 1.414 | 11.69 | 11.30 | 7.99 | 5.2392 ± 0.0023 | 6.51 | 84.7 | 2.50 | | | |
| 0.75 | F | 1.408 | 10.97 | 9.05 | 6.40 | 5.2434 ± 0.0002 | 6.55 | 84.2 | 2.32 | | | |
| 1 | F | 1.402 | 10.15 | 5.82 | 4.12 | 5.2425 ± 0.0004 | 7.32 | 92.2 | 2.06 | | | |

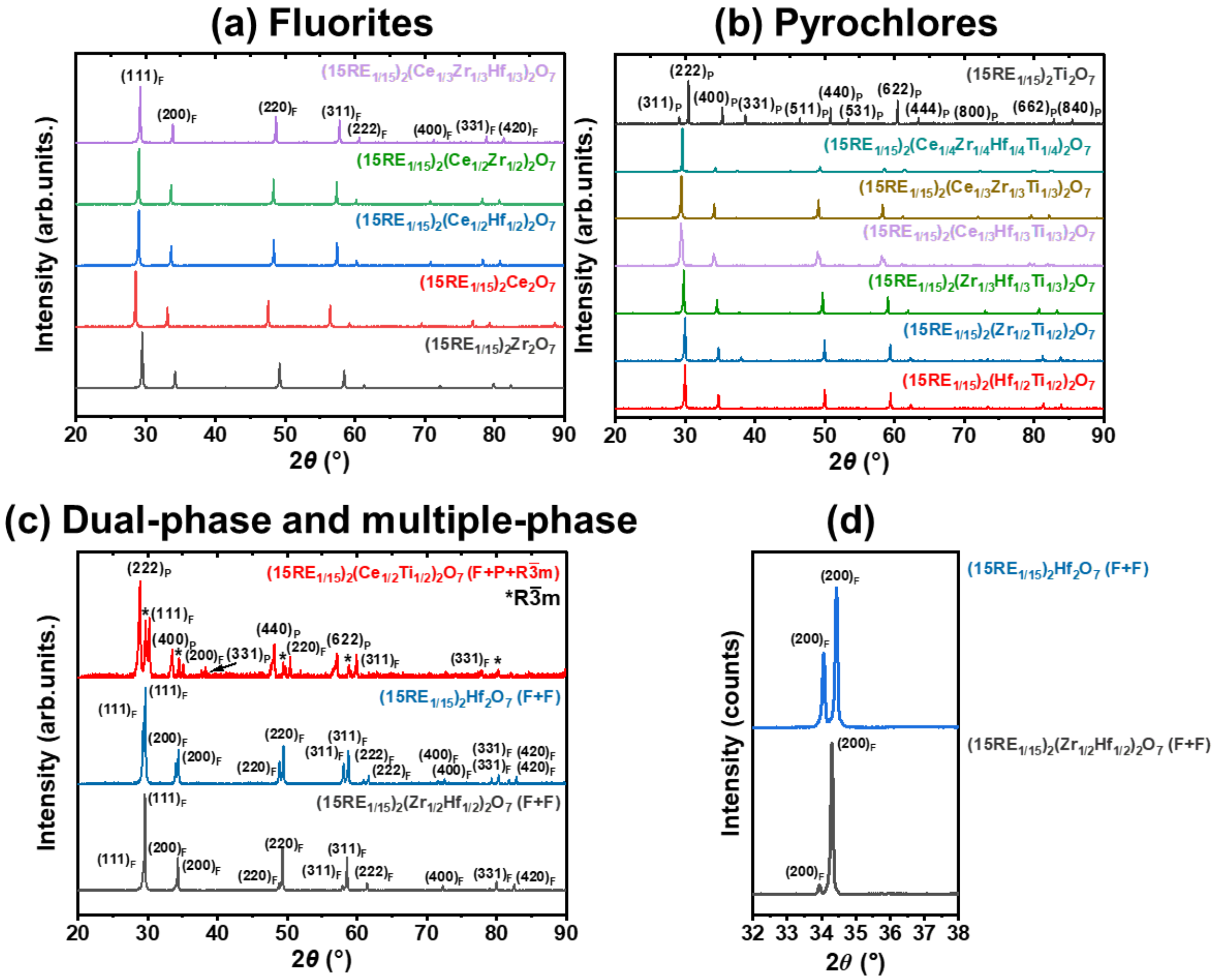

**Figure 1.** XRD patterns of the first group of 15 ultrahigh-entropy $(15RE_{1/15})_2TM_2O_7$ CCCs. The diffraction patterns are shown separately for (a) five single-phase fluorite compositions, (b) seven pyrochlore compositions, and (c) three dual- or multiphase compositions. (d) Slow step-scan XRD patterns in the 2θ range of 32–38° for the two fluorite–fluorite dual-phase compositions, $(15RE_{1/15})_2Hf_2O_7$ and $(15RE_{1/15})_2(Zr_{1/2}Hf_{1/2})_2O_7$ (denoted as "F+F"), clearly showing the coexistence of two distinct fluorite peaks and the absence of any detectable reflection near $2\theta \approx 37°$, which corresponds to the characteristic pyrochlore $(331)_P$ superstructure peak.

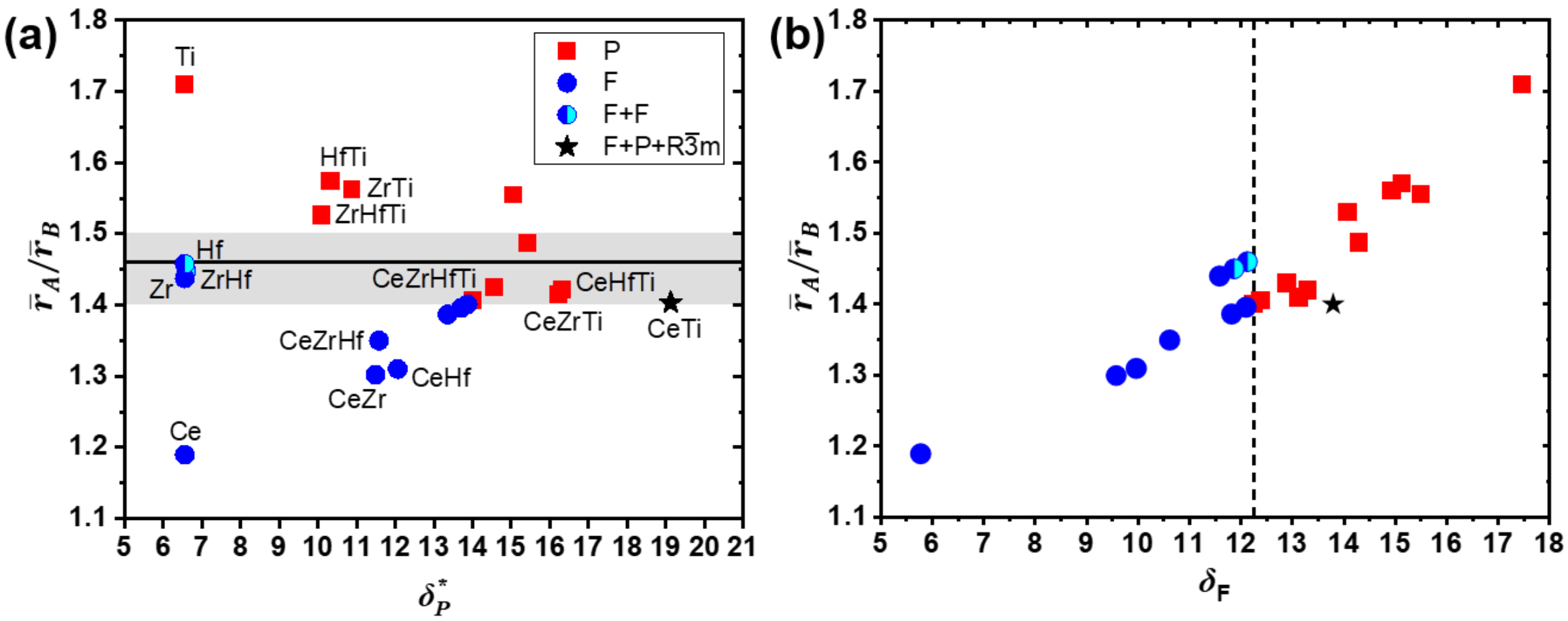

**Figure 2.** Phase stability of the first group of 15 ultrahigh-entropy $(15RE_{1/15})_2TM_2O_7$ CCCs and the second 19CCC-P|F$_y$ series, $(15RE_{1/15})_2[(Ce_{1/3}Zr_{1/3}Hf_{1/3})_yTi_{1-y}]_2O_7$ $(0 \leq y \leq 1)$, both with the ideal 2:2 ratio of A-type (larger 3+ rare-earth cations) and B-type (smaller 4+ cations), plotted in (a) $r_A/r_B$ *vs.* $\delta_P^*$ and (b) $r_A/r_B$ *vs.* $\delta_F$ spaces. The typical fluorite–pyrochlore transition threshold $(r_A/r_B = 1.46)$ for ternary $A_2B_2O_7$ systems is indicated by the black lines. The gray-shaded region in (a) represents the fluorite–pyrochlore dual-phase stability range $(1.40 \leq r_A/r_B \leq 1.50, \delta_P^* > 5)$ previously proposed for high-entropy $A_2B_2O_7$ compositions [refs]. The vertical dashed line in (b) represents a $\delta_F$ threshold used as an alternative criterion for the fluorite–pyrochlore transition.

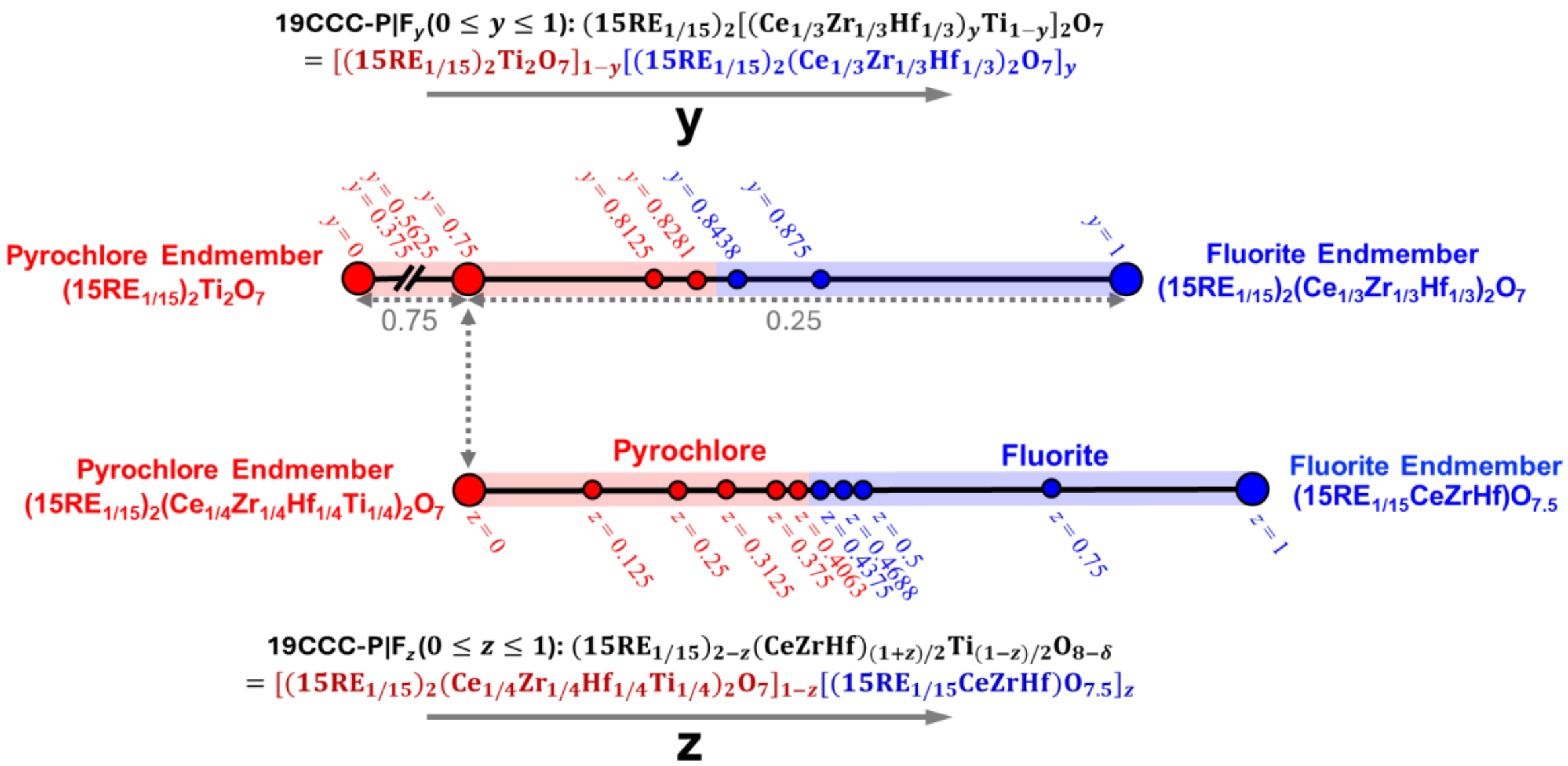

**Figure 3.** Schematic illustration of the pyrochlore-fluorite order-disorder transitions (ODTs) in the stoichiometric 19CCC-P|F$_y$ series, $(15RE_{1/15})_2[(Ce_{1/3}Zr_{1/3}Hf_{1/3})_yTi_{1-y}]_2O_7(0 \leq y \leq 1)$, and the non-stoichiometric 19CCC-P|F$_z$ series, $(15RE_{1/15})_{2-z}(CeZrHf)_{(1+z)/2}Ti_{(1-z)/2}O_{8-\delta}$ $(0 \leq z \leq 1)$, where $y$ or $z$ represents the fraction of the disordered fluorite endmember in each case. All 20 fabricated and characterized compositions exhibit single ultrahigh-entropy phases, with abrupt pyrochlore–fluorite ODTs occurring near $y \approx 0.83 - 0.84$ in the 19CCC-P|F$_y$ series and $z \approx 0.41 - 0.44$ in the 19CCC-P|F$_z$ series. Due to space constraints and for clarity of presentation, the 19CCC-P|F$_y$ series is not plotted to scale; the range from $y$ = 0 to $y$ = 0.75 has been compressed. Note that the composition $y$ = 0.75 in the 19CCC-P|F$_y$ series corresponds to $z$ = 0 in the 19CCC-P|F$_z$ series.

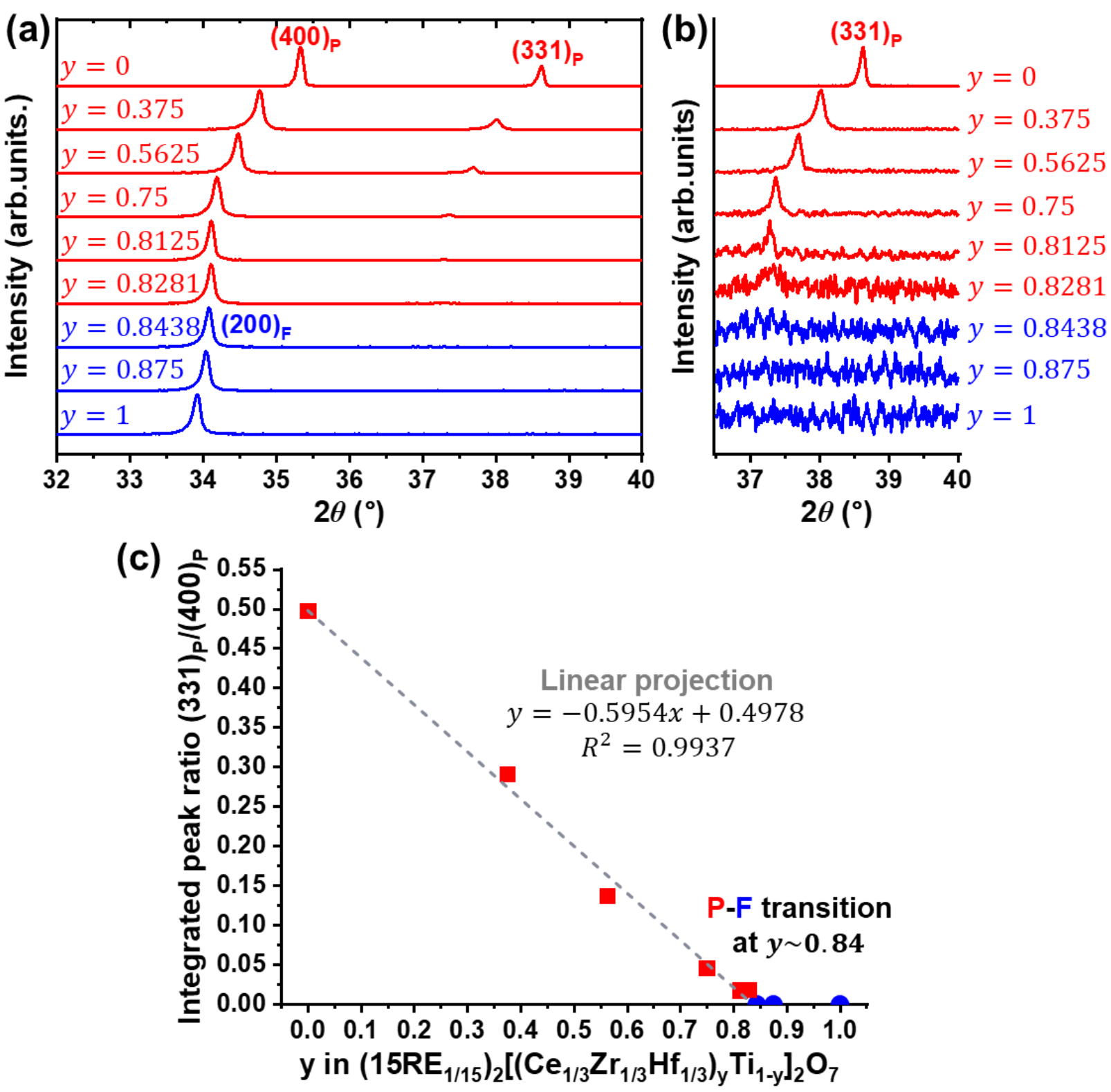

**Figure 4.** (a) Evolution of XRD patterns showing the fluorite–pyrochlore transition in the 19CCC-P|F$_y$ series, $[(15\mathrm{RE}_{1/15})_2\mathrm{Ti}_2\mathrm{O}_7]_{1-y}[(15\mathrm{RE}_{1/15})_2(\mathrm{Ce}_{1/3}\mathrm{Zr}_{1/3}\mathrm{Hf}_{1/3})_2\mathrm{O}_7]_y$ , where $y$ $(0 \leq y \leq 1)$ represents the fraction of the disordered defect-fluorite endmember, which can also be expressed in a simplified form as $(15\mathrm{RE}_{1/15})_2[(\mathrm{Ce}_{1/3}\mathrm{Zr}_{1/3}\mathrm{Hf}_{1/3})_y\mathrm{Ti}_{1-y}]_2\mathrm{O}_7$. (b) Intensity plots showing the disappearance of the characteristic pyrochlore superstructure peak $(331)_\mathrm{P}$ occurring between $y = 0.83$ and $y = 0.84$. (c) Normalized intensity of the characteristic pyrochlore superstructure peak $(331)_\mathrm{P}$, relative to the $(400)_\mathrm{P}$ peak, as a function of the compositional variable $y$, revealing the fluorite–pyrochlore ODT consistent with a linear projection of the superstructure peak intensity.

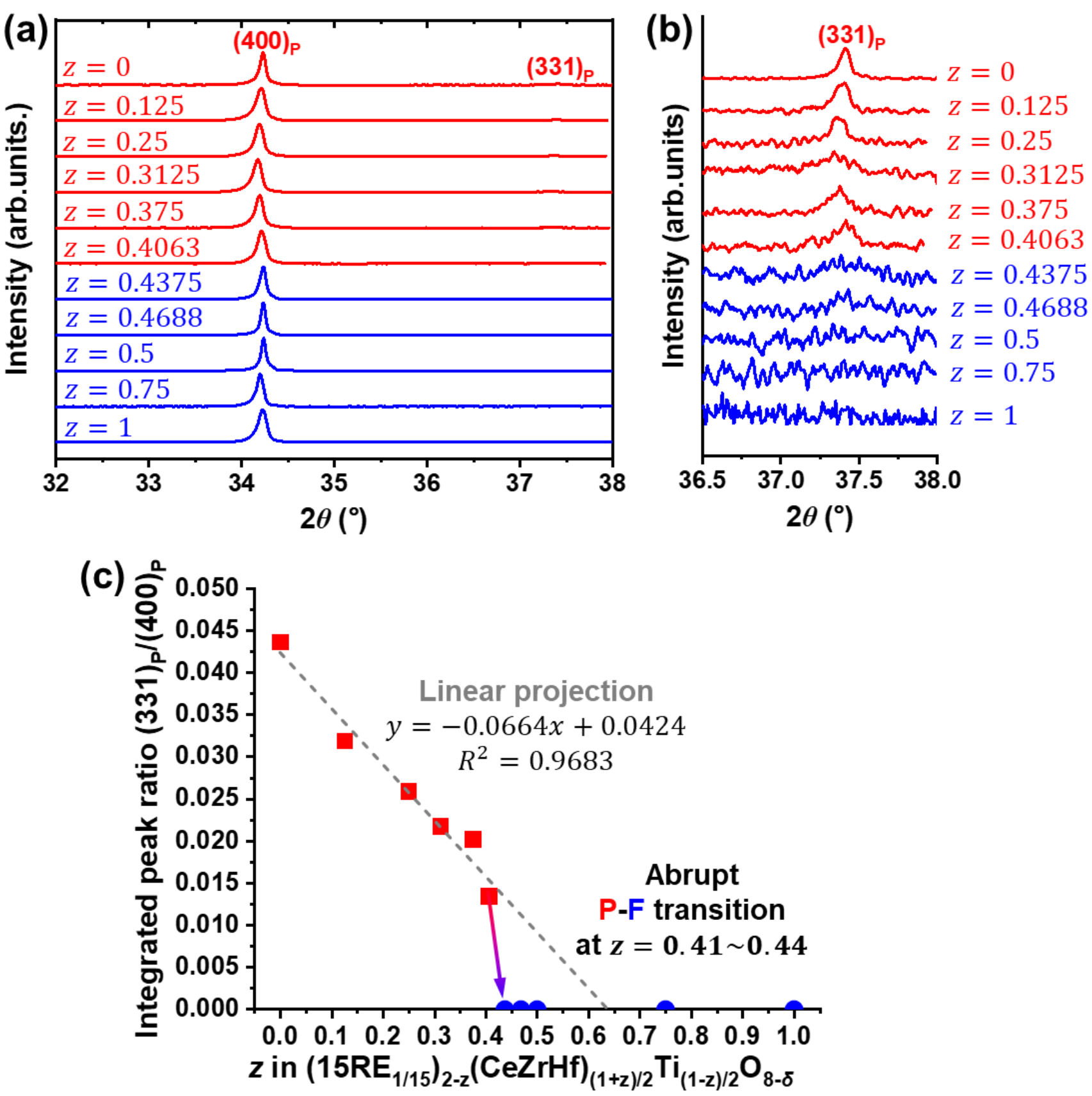

**Figure 5.** (a) Evolution of XRD patterns showing the fluorite–pyrochlore transition in the 19CCC-P|F$_z$ series, $[(15\mathrm{RE}_{1/15})_2(\mathrm{Ce}_{1/4}\mathrm{Zr}_{1/4}\mathrm{Hf}_{1/4}\mathrm{Ti}_{1/4})_2\mathrm{O}_7]_{1-z}[(15\mathrm{RE}_{1/15}\mathrm{CeZrHf})\mathrm{O}_{7.5}]_z$, where $z$ ($0 \leq z \leq 1$) represents the fraction of the disordered fluorite endmember, which can also be expressed in a simplified form as $(15\mathrm{RE}_{1/15})_{2-z}(\mathrm{CeZrHf})_{(1+z)/2}\mathrm{Ti}_{(1-z)/2}\mathrm{O}_{8-\delta}$. (b) Intensity plots showing the disappearance of the characteristic pyrochlore superstructure peak $(331)_\mathrm{P}$ between $z = 0.41$ and $z = 0.44$. (c) Normalized intensity of the characteristic pyrochlore superstructure peak $(331)_\mathrm{P}$, relative to the $(400)_\mathrm{P}$ peak, as a function of the compositional variable $z$. A linear projection of the peak intensity suggests a pyrochlore-to-fluorite ODT at $z \approx 0.64$, while the actual ODT occurs more abruptly between $z = 0.41$ and $z = 0.44$.

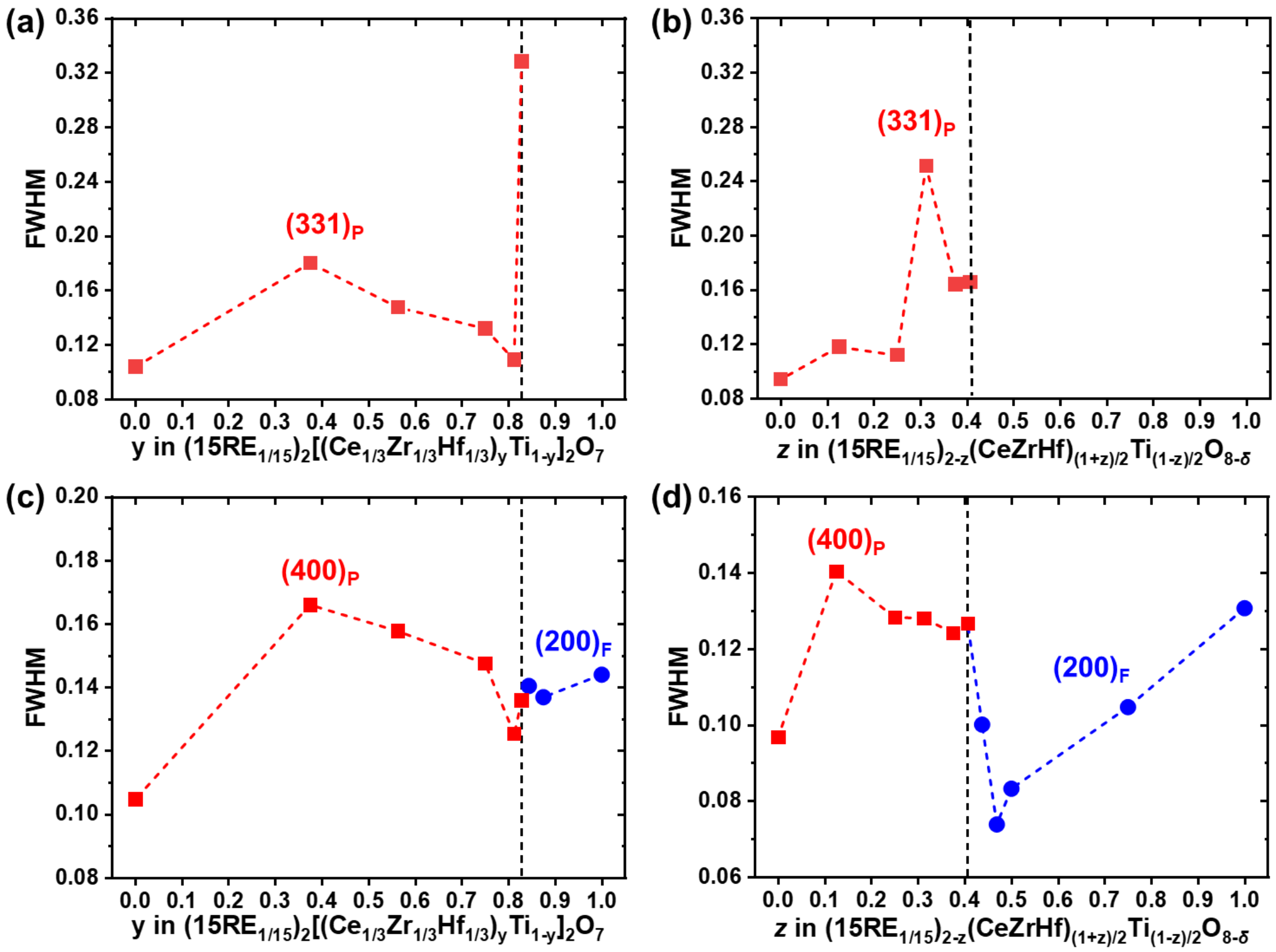

**Figure 6.** Evolution of the full width at half maximum (FWHM) of the pyrochlore $(331)_\mathrm{P}$ and $(400)_\mathrm{P}$ peaks and the fluorite $(200)_\mathrm{F}$ peak as a function of the compositional variable $y$ for (a, c) the stoichiometric 19CCC-P|F$_y$ series, $(15\mathrm{RE}_{1/15})_2[(\mathrm{Ce}_{1/3}\mathrm{Zr}_{1/3}\mathrm{Hf}_{1/3})_y\mathrm{Ti}_{1-y}]_2\mathrm{O}_7$ $(0 \leq y \leq 1)$, and as a function of the compositional variable $z$ for (b, d) the non-stoichiometric 19CCC-P|F$_z$ series, $(15\mathrm{RE}_{1/15})_{2-z}(\mathrm{CeZrHf})_{(1+z)/2}\mathrm{Ti}_{(1-z)/2}\mathrm{O}_{8-\delta}$ $(0 \leq z \leq 1)$. The gray vertical lines indicate the fluorite–pyrochlore ODTs.

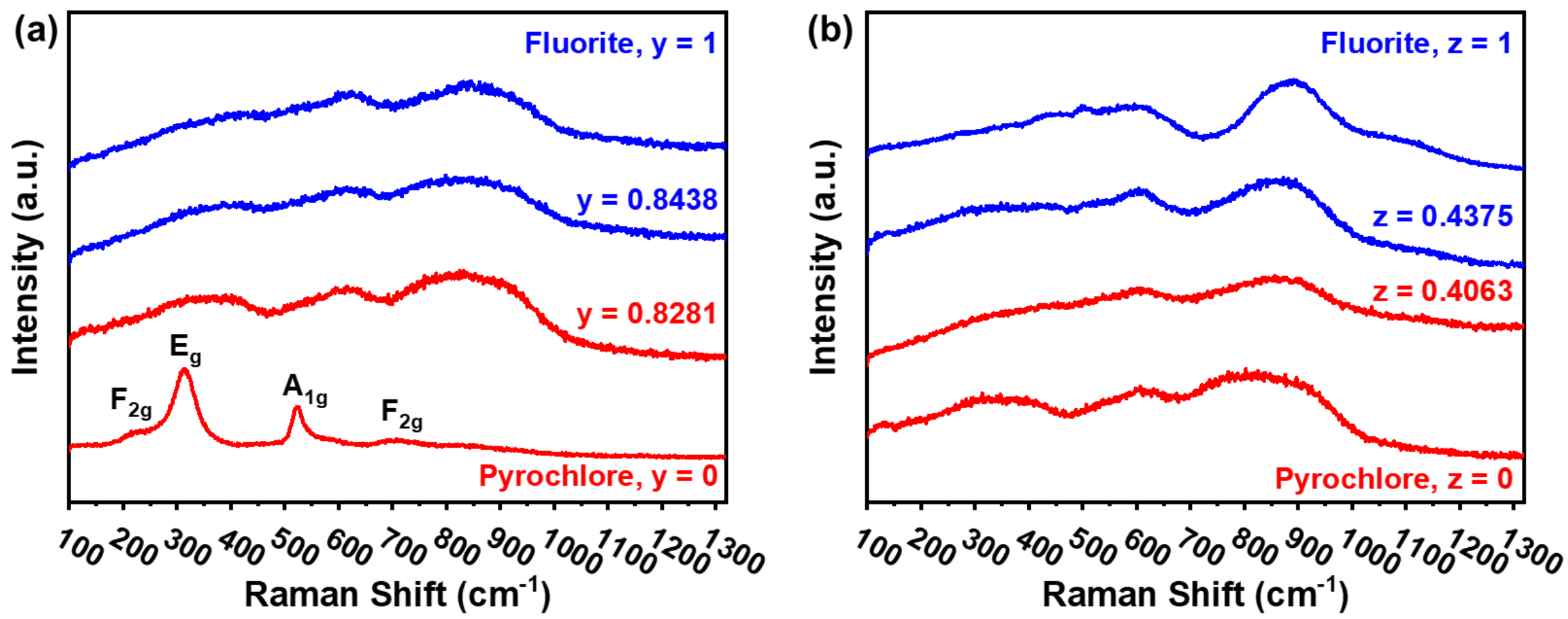

**Figure 7.** Raman spectra of selected compositions from two 19-component CCCs series exhibiting pyrochlore-fluorite ODTs: (a) the stoichiometric 19CCC-F|P$_y$ series, $(15\mathrm{RE}_{1/15})_2[(\mathrm{Ce}_{1/3}\mathrm{Zr}_{1/3}\mathrm{Hf}_{1/3})_{1-y}\mathrm{Ti}_y]_2\mathrm{O}_7$ ($0 \leq y \leq 1$), and (b) the non-stoichiometric 19CCC-F|P$_z$ series, $(15\mathrm{RE}_{1/15})_{2-z}(\mathrm{CeZrHf})_{(1+z)/2}\mathrm{Ti}_{(1-z)/2}\mathrm{O}_{8-\delta}$ ($0 \leq z \leq 1$). The selected compositions include the fluorite and pyrochlore endmembers ($y = 0,1;\ z = 0,1$) and compositions immediately before and after the pyrochlore-fluorite ODTs ($y = 0.8281,\ 0.8438;\ z = 0.4063, 0.4375$).

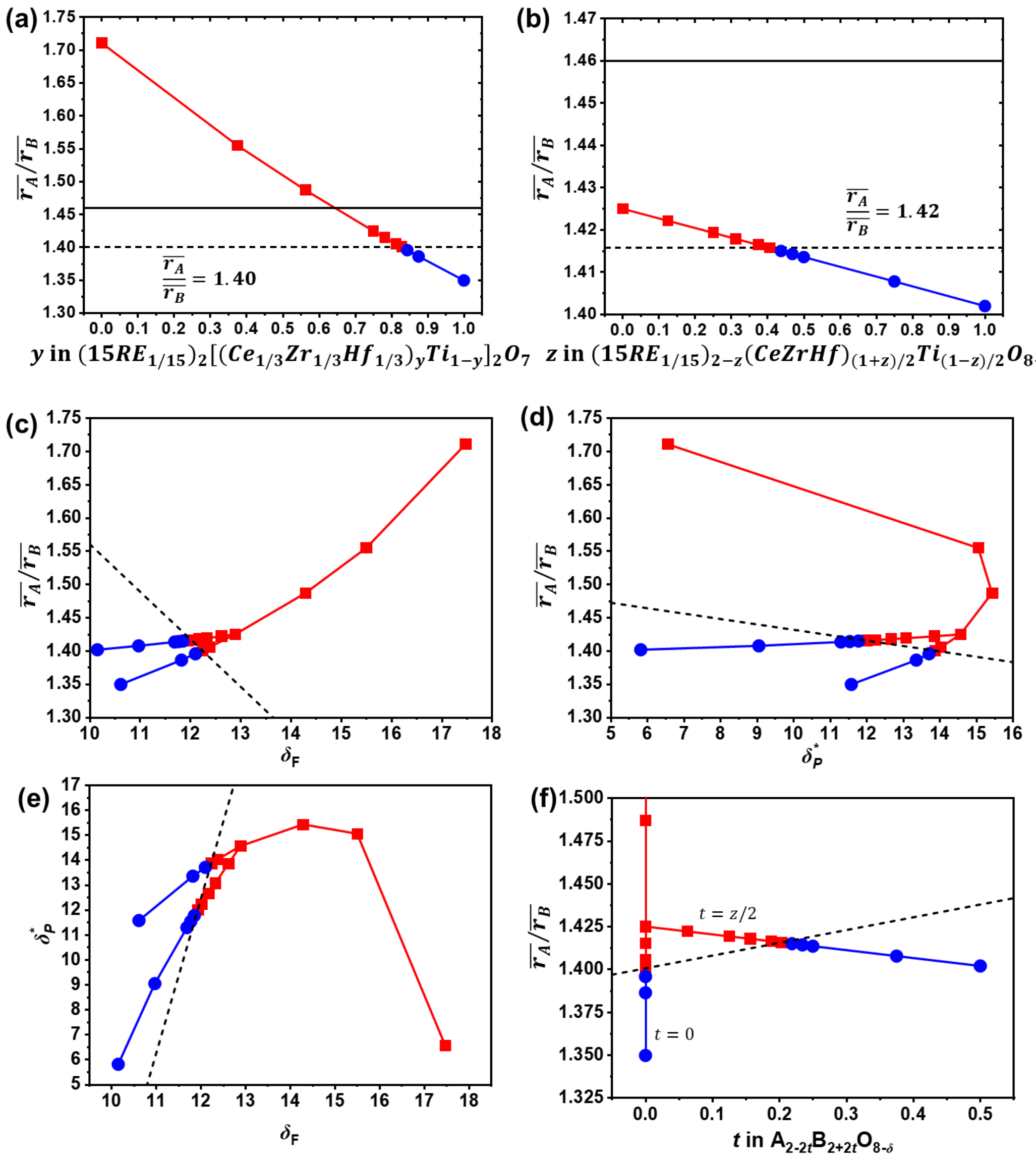

**Figure 8.** Averaged cation-radius ratio $\overline{r_A}/\overline{r_B}$ *vs.* (a) compositional variable $y$ for the stoichiometric 19CCC-P|F$_y$ series, $(15RE_{1/15})_2[(Ce_{1/3}Zr_{1/3}Hf_{1/3})_{1-y}Ti_y]_2O_7$ $(0 \leq y \leq 1)$, (b) compositional variable $z$ for the non-stoichiometric 19CCC-F|P$_z$ series, $(15RE_{1/15})_{2-z}(CeZrHf)_{(1+z)/2}Ti_{(1-z)/2}O_{8-\delta}$ $(0 \leq z \leq 1)$. Panels (c) and (d) show $\overline{r_A}/\overline{r_B}$ *vs.* size disorder $\delta_F$ and the modified size disorder $\delta_P^*$, respectively, for both 19CCC-P|F$_y$ and 19CCC-F|P$_z$ series. (e) $\delta_P^*$ *vs.* $\delta_F$ for both series. (f) $\overline{r_A}/\overline{r_B}$ *vs.* $(1-t)/(1+t)$, where t = 0 for the 19CCC-P|F$_y$ series and $t = z/2$ for the 19CCC-F|P$_z$ series.

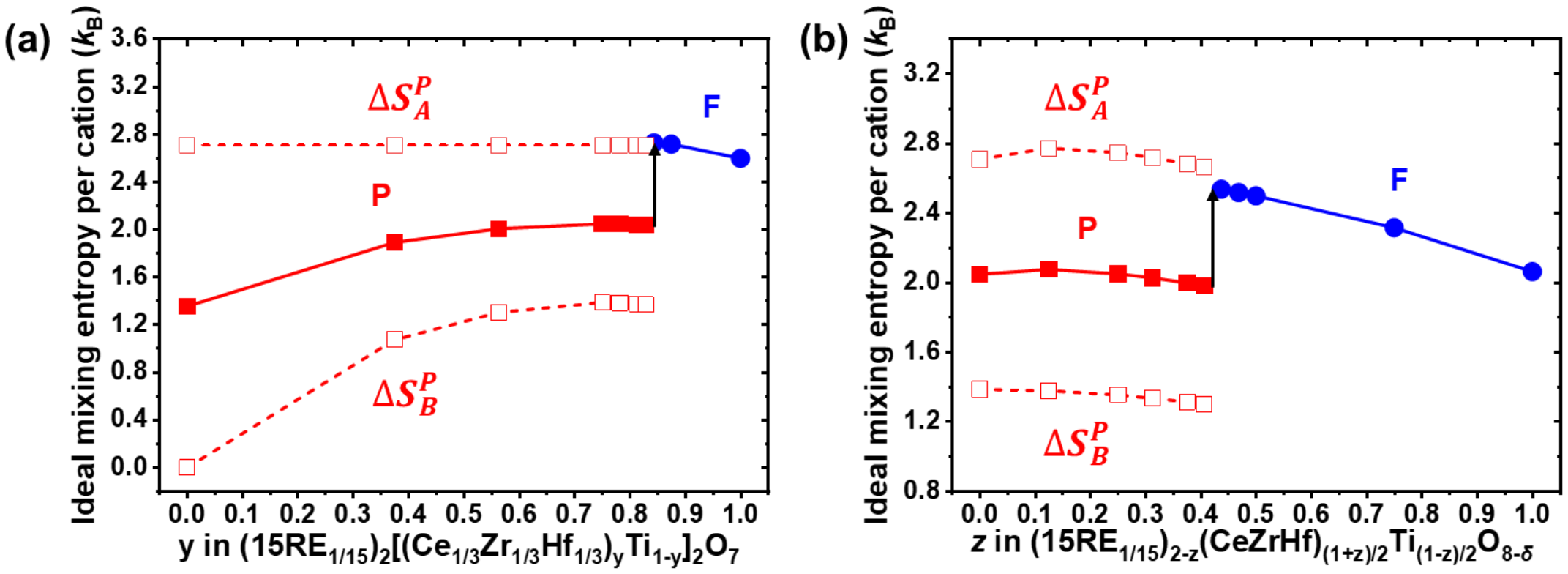

**Figure 9.** Computed ideal configurational entropy of mixing (in units of $k_B$ per cation) as a function of the compositional variables (a) $y$ for the 19CCC-P|F$_y$ series, $(15RE_{1/15})_2[(Ce_{1/3}Zr_{1/3}Hf_{1/3})_yTi_{1-y}]_2O_7$, and (b) $z$ for the 19CCC-P|F$_z$ series, $(15RE_{1/15})_{2-z}(Ce_{1/3}Zr_{1/3}Hf_{1/3})_{(3+3z)/2}Ti_{(1-z)/2}O_{8-\delta}$ . For the pyrochlore phase with two sublattices, the solid data points and solid lines represent the average configurational entropy, whereas the hollow data points and dashed lines represent the configurational entropies of the individual sublattices.

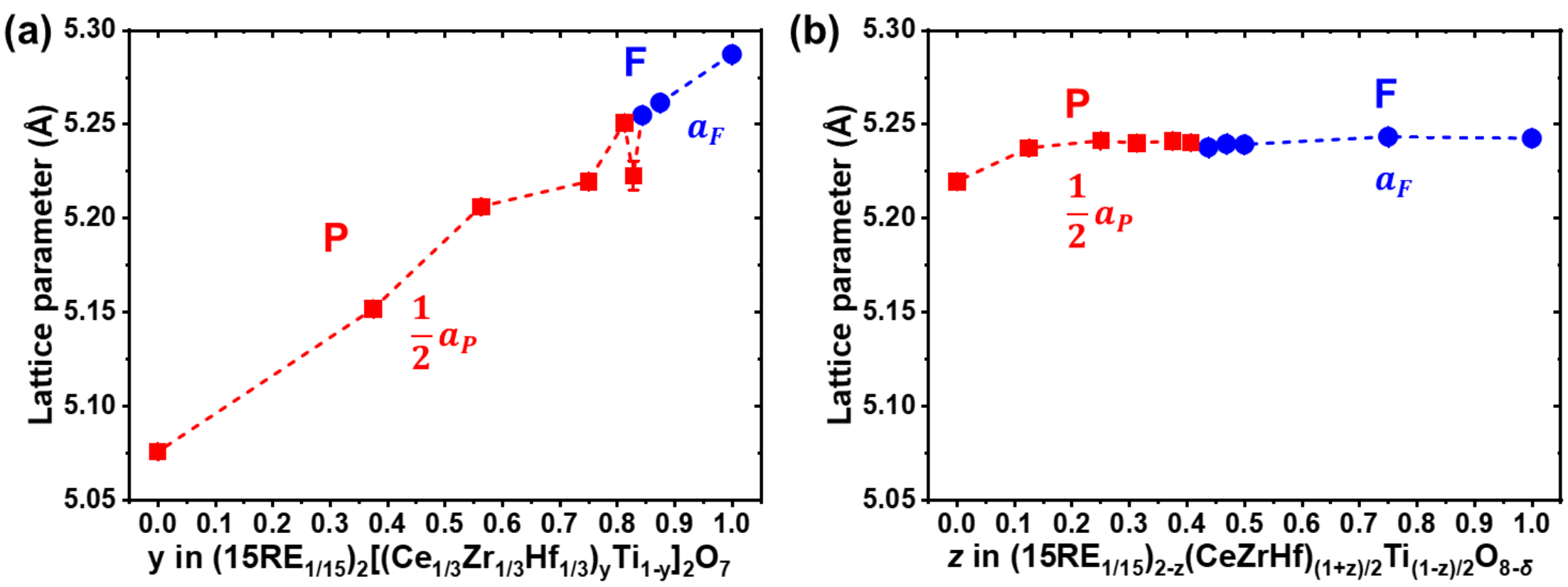

**Figure 10.** Measured lattice parameters ($a_{\mathrm{F}}$ for the fluorite phase and $a_{\mathrm{P}}/2$ for the pyrochlore phase) for (a) the 19CCC-P|F$_y$ series ($0 \leq y \leq 1$) and (b) the 19CCC-P|F$_z$ series ($0 \leq z \leq 1$). The lattice parameters were obtained from unit-cell refinements of the XRD patterns and normalized to the primitive fluorite cell parameter ($a_{\mathrm{F}}$) for direct comparison.